# A Method for Direct Estimation of Left Ventricular Global Longitudinal Strain Rate from Echocardiograms

**Authors:**

Brett A. Meyers[1], Melissa C. Brindise[1], Shelby Kutty[3], Pavlos P. Vlachos[1,2]*

**Affiliations:**

[1]School of Mechanical Engineering, Purdue University, 585 Purdue Mall, West Lafayette, IN 47907, USA

[2]Weldon School of Biomedical Engineering, Purdue University, 206 S. Martin Jischke Dr., West Lafayette, IN 47907, USA

[3]Taussig Heart Center, Johns Hopkins University School of Medicine, Baltimore, MD, 21287, USA

*Correspondence to Pavlos P. Vlachos, School of Mechanical Engineering, Purdue University, pvlachos@purdue.edu

Total word count: 4574

# Abstract

We present a new method for measuring global longitudinal strain and global longitudinal strain rate from 2D echocardiograms using a logarithmic-transform correlation (LTC) method. Traditional echocardiography strain analysis depends on user inputs and chamber segmentation, which yield increased measurement variability. In contrast, our approach is automated and does not require cardiac chamber segmentation and regularization, thus eliminating these issues. The algorithm was benchmarked against two conventional strain analysis methods using synthetic left ventricle ultrasound images. Measurement error was assessed as a function of contrast-to-noise ratio (CNR) using mean absolute error and root-mean-square error. LTC showed better agreement to the ground truth for strain ($R^2 = 0.91$) and strain rate ($R^2 = 0.85$) as compared to conventional algorithms (strain ($R^2 = 0.7$), strain rate ($R^2 = 0.7$)) and was unaffected by CNR. A 200% increase in strain measurement accuracy was observed compared to the conventional algorithms. Subsequently, we tested the method using a 54-subject clinical cohort (20 subjects diseased with cardiomyopathy, 34 healthy controls). Our method distinguished between normal and abnormal left ventricular function with an AUC = 0.85, a 10% improvement over the conventional GLS algorithms.

## Key terms

Echocardiography, global longitudinal strain, image registration, speckle tracking.

# Introduction

Global longitudinal strain (GLS), average strain of the cardiac chamber wall measured using speckle tracking echocardiography (STE), is used for the quantification of left ventricle (LV) function. Because GLS is more robust to reader error than LV ejection fraction[1,2], it is increasingly used more in clinical practice[3], especially for the detection of systolic dysfunction[4]. Moreover, the global longitudinal strain rate (GLSr), which is computed by differentiating GLS temporally, quantifies the rate of LV contraction and relaxation, thus providing information on systolic and diastolic function.

Limitations of STE can impact GLS and GLSr measurement accuracy. In order to perform a GLS-STE measurement, the user must provide an initial shape model of the LV boundary, thus introducing variability and reducing measurement reproducibility[5]. Boundary tracking is subsequently performed using block-matching or cross-correlation kernels[6,7], sensitive to image quality, spatial and temporal resolution, and signal dropout[8]. Furthermore, commercial tools use proprietary tracking and post-processing algorithms, making cross-platform comparison impractical[5,9]. Collectively these limitations hinder wide-spread acceptance of GLS and GLSr as diagnostic parameters with established standard ranges for normal and abnormal function[1,4,10–12].

Here we present a novel algorithm for direct GLS estimation from echocardiograms that overcomes the issues mentioned above. Our approach does not require LV boundary initialization, regional smoothing, or assumptions of LV shape. The algorithm directly measures the GLSr of the entire LV, which is then integrated in time to provide the GLS. Hence, this method is user-independent and robust to noise and image artifacts. Error analysis was performed using synthetic ultrasound images[13] and clinical demonstration was performed with patient data from healthy and cardiomyopathy subjects. In both cases, we compared the results from our algorithm against conventional STE algorithms.

# Materials and Methods

## Theory

During each heartbeat, the LV undergoes complex, three-dimensional motion as it contracts and relaxes. This motion is composed of a planar translation and deformation that relate how positions $(x_n, y_n, z_n)$ along the LV move to $(x_{n+1}, y_{n+1}, z_{n+1})$[14],

$$\begin{bmatrix} x_{n+1} \\ y_{n+1} \\ z_{n+1} \end{bmatrix} = \boldsymbol{F} \begin{bmatrix} x_n \\ y_n \\ z_n \end{bmatrix} + \vec{\boldsymbol{T}} = \begin{bmatrix} a_{11} & a_{12} & a_{13} \\ a_{21} & a_{22} & a_{23} \\ a_{31} & a_{32} & a_{33} \end{bmatrix} \begin{bmatrix} x_n \\ y_n \\ z_n \end{bmatrix} + \begin{bmatrix} t_1 \\ t_2 \\ t_3 \end{bmatrix}. \qquad\qquad 1$$

$\vec{\boldsymbol{T}}$ is the translation matrix, and $\boldsymbol{F}$ is the deformation gradient tensor, which is related to the displacement gradient tensor $\nabla\boldsymbol{u}$, as,

$$\boldsymbol{F} = \boldsymbol{I} - \nabla\boldsymbol{u}, \qquad\qquad 2$$

where $\boldsymbol{I}$ is the identity matrix.

Lagrange strain, $\boldsymbol{\varepsilon}$, is expressed as a function of $\nabla\boldsymbol{u}$ when motion and deformation are small, such that,

$$\boldsymbol{\varepsilon} = \frac{1}{2}(\nabla\boldsymbol{u} + \nabla\boldsymbol{u}^T). \qquad\qquad 3$$

Equation 3 can be written as the Lagrange strain equation,

$$\varepsilon = \frac{l - l_0}{l_0}, \qquad\qquad 4$$

where $l_0$ is the reference length and $l$ is the deformed length. The quantity $\varepsilon$ is the accepted definition of GLS[9]. As a result, GLS is a function of the deformation gradient tensor $\boldsymbol{F}$. In the following sections, we describe how we estimate GLS from the cross-correlation of two consecutive images.

**Pairwise Cross-Correlation**

Image cross-correlation provides a statistical estimate of the translation of an image pattern between two frames. This method is used in image registration[15], speckle tracking[16], particle image velocimetry[17], and image correlation[18]. The 2D discrete spatial cross-correlation between two images, $I_n$ and $I_{n+1}$, is expressed as,

$$R(x,y) = \sum_{i=-N/2}^{N/2} \sum_{j=-M/2}^{M/2} I_n(i,j) I_{n+1}(x+i, y+j), \qquad\qquad 5$$

where $R(x,y)$ is the correlation plane, $(i,j)$ are the summation indices of the correlation, and $N$ and $M$ are the image height and width, respectively. The cross-correlation can be performed in the spectral domain, as,

$$R(x,y) = \mathcal{F}^{-1}\big[\bar{\mathcal{F}}\big(I_n(x,y)\big)\mathcal{F}\big(I_{n+1}(x,y)\big)\big], \qquad\qquad 6$$

where $\mathcal{F}$ is the 2D Fourier transformation (FT) and $\bar{\mathcal{F}}$ is the complex conjugate of the FT. The expanded form of Equation 6 is written as,

$$R(x,y) = \iint \bar{G}_n(u,v) G_{n+1}(u,v) e^{-j2\pi(u(t_1+x)+v(t_2+y))} du\, dv, \qquad\qquad 7$$

where $(u,v)$ are wavenumbers proportional to positions $(x,y)$ and $G$ is the image FT.

The Fourier transform affine theorem stipulates that rotation, stretch, and shear occurs on the FT magnitude and phase[19]. We use this to establish how the affine transform affects the rigid translation estimate by replacing $G_{n+1}$ in Equation 7 with the relationship for $G_n$ using Equation 1,

$$R(x,y) = \iint \frac{\bar{G}_n(u,v)\, G_n(u',v')}{|F|} e^{-j2\pi\left(u'(t_1+x)+v'(t_2+y)\right)} du\, dv. \qquad\qquad 8$$

Here, $u' = a_{11}u + a_{21}v$, $v' = a_{12}u + a_{22}v$, and $|F| = det(F)$. Equation 8 provides a correlation plane with the peak shifted from the plane center by $(\Delta x, \Delta y)$, directly related to the translations $t_1$ and $t_2$ and the local deformation gradient tensor $F$. The correlation peak shifts are written as,

$$\Delta x \cong a_{11}t_1 + a_{12}t_2, \quad \Delta y \cong a_{21}t_1 + a_{22}t_2. \qquad\qquad 9$$

The deformation gradients $a_{ij}$ produce the strain captured by the GLS measurement.

## Translation-Invariant FT Magnitude Correlation for GLS Estimation

The components of $F$ can be estimated separately of $\vec{T}$ using the magnitude, $|F(u,v)|$, of the FT[19,20]. The cross-correlation of the FT magnitudes is *translation invariant* yielding the terms of $F$ with no contribution from $\vec{T}$. The *log-polar* basis Fourier-Mellin transform[21,22], popular in image registration, decouples terms of $F$ to estimate image rotation and stretch.

Contraction and relaxation of the LV result in deformation akin to anisotropic image rescaling. By changing the FT magnitude image coordinates from cartesian $(u,v)$ to orthogonal logarithmic coordinates $(log u, log v)$, the resulting displacements from the correlation between two FT magnitude images now correspond to horizontal $\Delta u$ and vertical $\Delta u$ rescaling such that,

$$G(log u + log \Delta u, log v + log \Delta v). \qquad\qquad 10$$

This correlation is affected by rotation and shear based on the terms present in Equation 8. However, if those terms are minimized beforehand, the correlation peak shift estimates $(\Delta u, \Delta v)$ can be related to the terms $a_{11}$ and $a_{22}$ from $F$ through,

$$a_{11} = \Delta u \cong e^{\Delta x}, \quad a_{22} = \Delta v \cong e^{\Delta y}. \qquad\qquad 11$$

We substitute $a_{11}$ and $a_{22}$ from Equation 11 into **F** in Equation 2 and solve for $\nabla\mathbf{u}$ such that,

$$\nabla\mathbf{u} = \begin{bmatrix} e^{dx} - 1 & 0 \\ 0 & e^{dy} - 1 \end{bmatrix}. \qquad\qquad 12$$

Since deformation of the LV in long axis apical (ALAX) scans occurs along its length, we assume GLS occurs predominantly along the vertical direction, providing the GLS estimator,

$$\varepsilon_{GLS} = \frac{l - l_0}{l_0} \approx e^{dy} - 1. \qquad\qquad 13$$

**Direct Global Longitudinal Strain Estimation Algorithm**

We now describe our algorithm using the translation-invariant FT magnitude correlation to estimate the GLS based on Equation 13. A schematic of our algorithm is provided in Figure 1. The algorithm comprises two stages – the first performs an image registration to minimize shear and rotation that corrupt the correlation accuracy, while the second performs the GLS estimation.

The algorithm begins by selecting frames to analyze (Figure 1a). Next, the user selects three points from the first frame (Figure 1b), corresponding to approximate locations of the LV apex, the annulus septal, and annulus lateral positions. These points are tracked between consecutive frames using standard pairwise cross-correlation (Figure 1c). For each frame, the geometric center from the tracked points is computed along with the vertical axis's orientation angle for a line formed from the annulus center to the apex (Figure 1d). Frames are then aligned based on the geometric center, and the orientation angle is corrected. A circular ROI for each frame is defined from the tracked points.

In the second stage, for each pair of sequential registered images, *t* and *t+1* (Figure 1e), their FT and FT magnitude is computed. The FT magnitudes are interpolated from the image grid onto a logarithm-scale grid (Figure 1f). Because the FT logarithm transformed images are symmetric about the image diagonals, they are separated into four quarters to improve measurement accuracy. Each sub-image is filtered[23] and their FT is computed (Figure 1g). The FT sub-images are then correlated using the *spectral* cross-correlation kernel and ensemble-averaged to provide the displacements ($\Delta x, \Delta y$) (Figure 1h). A dynamic phase-filtered kernel is applied to the cross-correlation to improve estimate accuracy[24–26]. The displacements $(\Delta x, \Delta y)$ are adjusted based on the logarithm-scale grid, becoming $(\Delta x', \Delta y')$. The GLSr between frames is computed using $\Delta y'$ and Equation 13. Finally, GLSr across each frame pair is integrated in time using 4th-Order Runge-Kutta to obtain GLS (Figure 1i). The integral operator provides inherent smoothing, which suppresses noise in the GLS measurements. We will hereafter refer to this method as the Logarithm-Transform Correlation (LTC) method.

## Speckle Tracking Strain

This study uses two standard STE algorithms against which we benchmark our method. One algorithm uses the spatial cross-correlation kernel introduced in Equation 5[16], referred to herein as the *Direct Cross-Correlation* or DCC method. The second uses the spectral cross-correlation (Equation 6), hereafter referred to as the Fourier Transform Correlation or FTC method.

Boundary tracking was performed by propagating the segmented boundary of the initial frame through the estimated displacement fields using 4th-Order Runge-Kutta. GLS was estimated as the measured change in arc-length between the segmented and the tracked boundary from each frame. Image co-registration was not performed, as it is not required to obtain a consistent GLS measurement based on the arc-length change calculation.

## Artificial Echocardiograms

Error analysis was performed using synthetic LV ALAX echocardiograms[13,27]. Ground truth boundaries, displacements, and strains for each dataset were included with the synthetic images. Mean absolute error (MAE) and root-mean-square error (RMSE) for GLS and GLSr were quantified as a function of contrast to noise ratio (CNR). CNR was defined as the ratio of L2-norm of the tissue signal to L2-norm of the signal inside the LV[28]. Error quantities were normalized by the peak GLS or peak GLSr.

## Clinical Imaging

The method's clinical capabilities were demonstrated using a cohort of pediatric patients with confirmed cardiomyopathy and age-matched controls collected from a study conducted at the University of Nebraska Medical Center in Omaha, Nebraska, USA. The Institutional Review Boards of Purdue, Nebraska, and Johns Hopkins Universities each approved the study protocol. All procedures were performed in accordance with relevant guidelines and regulations. Informed written consent was obtained from study subject guardians for those under age 18 or from the subject themself for those over age 18.

Each patient underwent a routine echocardiogram study on an iE33 ultrasound system (Philips Healthcare, Andover, MA, USA). Studies were collected based on the American Society of Echocardiography recommendations[29]. The 54-subject cohort included 4 patients with confirmed dilated cardiomyopathy (DCM), 16 patients with confirmed hypertrophic cardiomyopathy (HCM), and 34 age-matched controls. Information on the cohort demographics is provided in Table 1 and heart function indices in Table 2.

Doppler measurements were collected in the ALAX A4C view. B-mode ALAX A2C, A3C, and A4C view scans were performed conventionally, not explicitly collected for strain measurements. Images were stored in Digital Imaging and Communications in Medicine (DICOM) format for post-processing. 900 individual heartbeats were analyzed, consisting of 573 control, 261 HCM, and 66 DCM. Peak absolute GLS, peak absolute systolic GLSr (GLSrs), and peak absolute early diastolic GLSr (GLSre) were computed. Ventricle dimension measurements were computed by the Simpson biplane method using the GE EchoPAC software. Data were classified into control (CTRL) and cardiomyopathy (CM) groups. Statistical significance was tested using Student's t-test between methods among the same condition and between each condition. Receiver operating characteristic (ROC) curves and area under the curve (AUC) were computed for each parameter.

## Results

### Error Analysis Results

Figure 2 presents error analysis results. The GLSr estimates (Figure 2a-1) by the DCC and FTC method yielded linear regression fits with similar slopes and bias of approximately $m \approx 0.7 \; s \cdot s^{-1}$ and $b \approx 0.025 \; s^{-1}$, and quality of fit of $R^2_{DCC}$=0.72 and $R^2_{FTC}$=0.76. Conversely, LTC GLSr estimates had a slope fit of $m = 0.92 \; s \cdot s^{-1}$ and bias of $b = 0.016 \; s^{-1}$ with a fit quality of $R^2_{LTC}$=0.85. GLSr, as a function of CNR, is shown in Figure 2a-2 (values are normalized by $GLSr = 0.95 \; s^{-1}$). LTC showed a 1.5 to 2-fold improvement in accuracy than DCC and a 1.5-fold improvement compared to FTC. Additionally, the LTC method is unaffected by CNR. In contrast, the FTC and DCC methods show a CNR dependence, albeit less for the FTC method.

GLS estimates by the DCC and FTC method maintained linear regression fits (Figure 2b-1) with similar slopes and biases of around $m \approx 0.6$ and $b \approx 0.3\%$, and quality of fit of $R_{DCC}^2$=0.71 and $R_{FTC}^2$=0.74. LTC GLS estimates show a slope of $m = 0.92$, a bias of $b = 0.09\%$, and fit quality of $R_{LTC}^2$=0.91. The LTC method shows more than 200% improvement in measurement accuracy compared against the DCC and FTC methods as a function of CNR (Figure 2b-2), where values are normalized by $GLS = 8.47\%$. The error analysis demonstrates that the LTC method is unaffected by CNR, while the DCC and FTC methods are affected by signal quality.

**Clinical Analysis Results**

Results comparing the LTC method against conventional GLS methods for each parameter are presented in Figure 3a-1 through 3c-1. LTC consistently measured increased peak GLS, GLSrs, and GLSre, compared to the conventional methods. FTC mean peak GLS measurements were increased compared to the DCC mean peak GLS. However, the FTC maintained an increased variance across both health states. This occurred for FTC GLSre and GLSrs, and the results are similar for the DCC method. Significance tests indicated the group means were statistically significant ($p < 0.05$) for all cases.

ROC curves are presented in Figure 3a-2 through 3c-2. The peak GLS ROC curves (Figure 3a-2) show that the LTC method AUC was 0.85, while the FTC and DCC AUCs were near 0.75. Peak GLSrs ROC curves (Figure 3b-2) show that all methods performed similarly in detecting abnormal function for this dataset. The LTC had $p < 0.05$ when compared to FTC but showed weak significance to DCC. Finally, for the peak GLSre ROC curves (Figure 3c-2), the LTC maintained an AUC of 0.82, and the FTC method had an AUC of 0.75, but both methods did not achieve the AUC from the peak GLS. Additionally, LTC achieved $p < 0.05$ against both methods.

# Discussion

This study presents a new algorithm, the LTC method, for computing GLS and GLSr estimates from ultrasound scans. Error analysis using synthetic ultrasound images demonstrated and quantified the LTC method's improvement over two current STE methods, and a clinical cohort was analyzed using the LTC and the two STE methods.

The LTC method does not rely on LV shape assumptions and avoids the use of boundary segmentation. Furthermore, because LV segmentation is not required, regularization to preserve the segmentation shape is avoided. Lastly, because the entire image of the LV is used to compute strain, out-of-plane motion is averaged out, making this method uniquely robust this error source.

The LTC method is novel by directly computing the GLSr between sequential frames, ensuring a reliable rate measurement. Computing GLS by integrating the GLSr, provides a smoothing operation that reduces noise. Commercial GLS algorithms have constraints that enforce tracked boundaries that are smooth in space and time, providing measurements that appear physically consistent. However, the returned measurements are also driven by the regularization. Thus, the GLS measurements do not adhere to the underlying deformation that occurs between frames.

The error analysis presented in Figure 2 demonstrated that the DCC and FTC methods are dependent on CNR. As CNR approaches 1, the mean pixel intensity of the LV wall and the mean pixel intensity of speckle noise are almost equal. This means the speckle noise cannot be differentiated from the LV wall, making the physical features ambiguous. Both methods underestimated the GLS, supporting the notion that it may be best to avoid such computation[5]. In contrast, the LTC method enables robust GLS computation even with noisy images, as supported by the MAE and RMSE plots in Figure 2.

Comparison of each method for all disease conditions, presented in Figure 3a-1, b-1, and c-1, provides the measurement distributions and their statistical significance. For each method, the CTRL and CM distributions overlapped while statistical significance was observed. The LTC method reported larger GLS and GLSr values compared to both STE methods. Between methods, statistical significance was observed while distributions overlapped. Comparison results suggest normal function will report larger GLS and GLSr values than abnormal function but establishing improvement of the LTC method through this analysis alone is not possible.

The ROC curves, presented in Figure 3a-2, 3b-2, and 3c-2, are used to determine if clinical separation is possible. The LTC method peak GLS and GLSre ROC curves show a 10% improvement in classification, which would improve the correct diagnosis rate.

LTC peak GLSre measurements agreed with the literature (peak GLSre > 1s$^{-1}$)[30], while GLS and GLSrs measurements were below nominal ranges from literature (peak GLS> 18%; peak GLSrs > 1s$^{-1}$)[31]. Commercial methods rely on regularization steps, which force the measurements to fit in LV shape models[5,9], thereby causing the GLS measurements to be a regularization function instead of the actual underlying deformation, feasibly leading to overestimation.

LTC algorithm limitations stem from the assumption that GLS and GLSr can be reliably measured from the septal and lateral walls, ignoring shortening near the apex. Furthermore, the algorithm assumes that reorientation has been performed correctly with no misalignment. If misalignment exists, the measurements will be a combination of GLS and off-axis strain. Finally, this study was limited by its clinical validation, which was performed using retrospective pediatric data. A dataset optimized for GLS estimation would help further substantiate findings. Additional testing should be performed using adult heart failure subjects.

We presented a new correlation kernel, the logarithm transform correlation or LTC, for quantifying GLS and GLSr from echocardiography scans. Our LTC-based algorithm does not use LV shape assumptions, is machine-agnostic, automated, and free of heuristic inputs. We compared the LTC against STE algorithms using artificial scans, analyzing error against ground truth GLS and GLSr values, and validated using clinical data from a study of pediatric cardiomyopathies. Results showed that the LTC method is unaffected by the image quality, providing improved measurement accuracy against the STE methods for both the synthetic data and clinical cohort.

## References

1.  Kalam, K., Otahal, P. & Marwick, T. H. Prognostic implications of global LV dysfunction: A systematic review and meta-analysis of global longitudinal strain and ejection fraction. *Heart* **100**, 1673–1680 (2014).

2.  Stanton, T., Leano, R. & Marwick, T. H. Prediction of all-cause mortality from global longitudinal speckle strain: comparison with ejection fraction and wall motion scoring. *Circ. Cardiovasc. Imaging* **2**, 356–364 (2009).

3.  Lang, R. M. *et al.* Recommendations for chamber quantification. *Eur. J. Echocardiogr.* **7**, 79–108 (2006).

4.    Smiseth, O. A., Torp, H., Opdahl, A., Haugaa, K. H. & Urheim, S. Myocardial strain imaging: how useful is it in clinical decision making? *Eur. Heart J.* **37**, 1196–1207 (2016).

5.    Lang, R. M. *et al.* Recommendations for Cardiac Chamber Quantification by Echocardiography in Adults: An Update from the American Society of Echocardiography and the European Association of Cardiovascular Imaging. *J. Am. Soc. Echocardiogr.* **28**, 1-39.e14 (2015).

6.    Amundsen, B. H. *et al.* Noninvasive Myocardial Strain Measurement by Speckle Tracking Echocardiography. *J. Am. Coll. Cardiol.* **47**, 789–793 (2006).

7.    Helle-Valle, T. *et al.* New noninvasive method for assessment of left ventricular rotation: speckle tracking echocardiography. *Circulation* **112**, 3149–3156 (2005).

8.    Rösner, A. *et al.* The influence of frame rate on two-dimensional speckle-tracking strain measurements: a study on silico-simulated models and images recorded in patients. *Eur. Hear. Journal-Cardiovascular Imaging* **16**, 1137–1147 (2015).

9.    Voigt, J. U. *et al.* Definitions for a common standard for 2D speckle tracking echocardiography: consensus document of the EACVI/ASE/Industry Task Force to standardize deformation imaging. *Eur. Hear. J. - Cardiovasc. Imaging* **16**, 1–11 (2015).

10.   Farsalinos, K. E. *et al.* Head-to-Head Comparison of Global Longitudinal Strain Measurements among Nine Different Vendors: The EACVI/ASE Inter-Vendor Comparison Study. *J. Am. Soc. Echocardiogr.* **28**, 1171-1181.e2 (2015).

11.   Yingchoncharoen, T., Agarwal, S., Popović, Z. B. & Marwick, T. H. Normal Ranges of Left Ventricular Strain: A Meta-Analysis. *J. Am. Soc. Echocardiogr.* **26**, 185–191 (2013).

12.   Menting, M. E. *et al.* Normal myocardial strain values using 2D speckle tracking echocardiography in healthy adults aged 20 to 72 years. *Echocardiography* **33**, 1665–1675 (2016).

13.   Alessandrini, M. *et al.* Realistic Vendor-Specific Synthetic Ultrasound Data for Quality Assurance of 2-D Speckle Tracking Echocardiography: Simulation Pipeline and Open Access Database. *IEEE Trans. Ultrason. Ferroelectr. Freq. Control* **65**, 411–422 (2018).

14. Meunier, J. & Bertrand, M. Echographic Image Mean Gray Level Changes with Tissue Dynamics: A System-Based Model Study. *IEEE Trans. Biomed. Eng.* **42**, 403–410 (1995).

15. Anuta, P. E. Spatial registration of multispectral and multitemporal digital imagery using fast fourier transform techniques. *IEEE Trans. Geosci. Electron.* **8**, 353–368 (1970).

16. Bohs, L. N. & Trahey, G. E. A Novel Method for Angle Independent Ultrasonic Imaging of Blood Flow and Tissue Motion. *IEEE Trans. Biomed. Eng.* **38**, 280–286 (1991).

17. Willert, C. E. & Gharib, M. Digital Particle Image Velocimetry. *Exp. Fluids* **10**, 181–193 (1991).

18. Chu, T. C., Ranson, W. F. & Sutton, M. a. Applications of Digital Image Correlation Techniques to Experimental Mechanics. *Exp. Mech.* **25**, 232–244 (1985).

19. Bracewell, R. N., Chang, K.-Y., Jha, A. K. & Wang, Y.-H. Affine theorem for two-dimensional Fourier transform. *Electron. Lett.* **29**, 304–304 (1993).

20. Reddy, B. S. & Chatterji, B. N. An FFT-based technique for translation, rotation, and scale-invariant image registration. *IEEE Trans. image Process.* **5**, 1266–1271 (1996).

21. Chen, Q., Defrise, M. & Deconinck, F. Symmetric Phase-Only Matched Filtering of Fourier-Mellin Transforms for Image Registration and Recognition. *IEEE Trans. Pattern Anal. Mach. Intell.* **16**, 1156–1168 (1994).

22. Giarra, M. N., Charonko, J. J. & Vlachos, P. P. Measurement of fluid rotation, dilation, and displacement in particle image velocimetry using a Fourier–Mellin cross-correlation. *Meas. Sci. Technol.* **26**, 35301 (2015).

23. Eckstein, A. & Vlachos, P. P. Assessment of advanced windowing techniques for digital particle image velocimetry (DPIV). *Meas. Sci. Technol.* **20**, 075402 (2009).

24. Meyers, B. A., Goergen, C. J. & Vlachos, P. P. Development and Validation of a Phase-Filtered Moving Ensemble Correlation for Echocardiographic Particle Image Velocimetry. *Ultrasound Med. Biol.* **44**, 477–488 (2018).

25. Eckstein, A. & Vlachos, P. P. Digital particle image velocimetry (DPIV) robust phase correlation. *Meas. Sci. Technol.* **20**, 055401 (2009).

26. Eckstein, A. C., Charonko, J. & Vlachos, P. Phase correlation processing for DPIV measurements. *Exp. Fluids* **45**, 485–500 (2008).

27. Alessandrini, M. *et al.* Generation of ultra-realistic synthetic echocardiographic sequences to facilitate standardization of deformation imaging. in *2015 IEEE 12th International Symposium on Biomedical Imaging (ISBI)* vols 2015-July 756–759 (IEEE, 2015).

28. Bell, M. A. L., Goswami, R., Kisslo, J. A., Dahl, J. J. & Trahey, G. E. Short-lag spatial coherence imaging of cardiac ultrasound data: Initial clinical results. *Ultrasound Med. Biol.* **39**, 1861–1874 (2013).

29. Lopez, L. *et al.* Recommendations for quantification methods during the performance of a pediatric echocardiogram: a report from the Pediatric Measurements Writing Group of the American Society of Echocardiography Pediatric and Congenital Heart Disease Council. *J. Am. Soc. Echocardiogr.* **23**, 465–495 (2010).

30. Morris, D. A. *et al.* Lower limit of normality and clinical relevance of left ventricular early diastolic strain rate for the detection of left ventricular diastolic dysfunction. *Eur. Hear. Journal-Cardiovascular Imaging* **19**, 905–915 (2017).

31. Jashari, H. *et al.* Normal ranges of left ventricular strain in children: a meta-analysis. *Cardiovasc. Ultrasound* **13**, 1021 (2015).

## Acknowledgments

The authors thank Mary J. Craft for her time and effort in data collection, aggregation, and sharing between universities. The support of CTSI and NIH (UL1TR002529) is gratefully acknowledged.

## Author Contributions

BAM, MCB, and PPV designed the algorithms. SK designed the study and coordinated subject enrollment. BAM performed the data analysis. BAM wrote the paper and all other authors made significant contributions to the writing.

## Funding

This project was supported by the Indiana Clinical and Translational Sciences Institute and funded by Grant Number UL1TR002529 from the National Institutes of Health,

National Center for Advancing Translational Sciences, Clinical, and Translational Sciences Award.

**Competing interests**

BAM and PPV are employees of Cordian Technologies, a company having a portfolio of patents for B-mode and Doppler echocardiogram-based measurements. MCB and SK have no potential conflict of interest.

**Figure Legends**

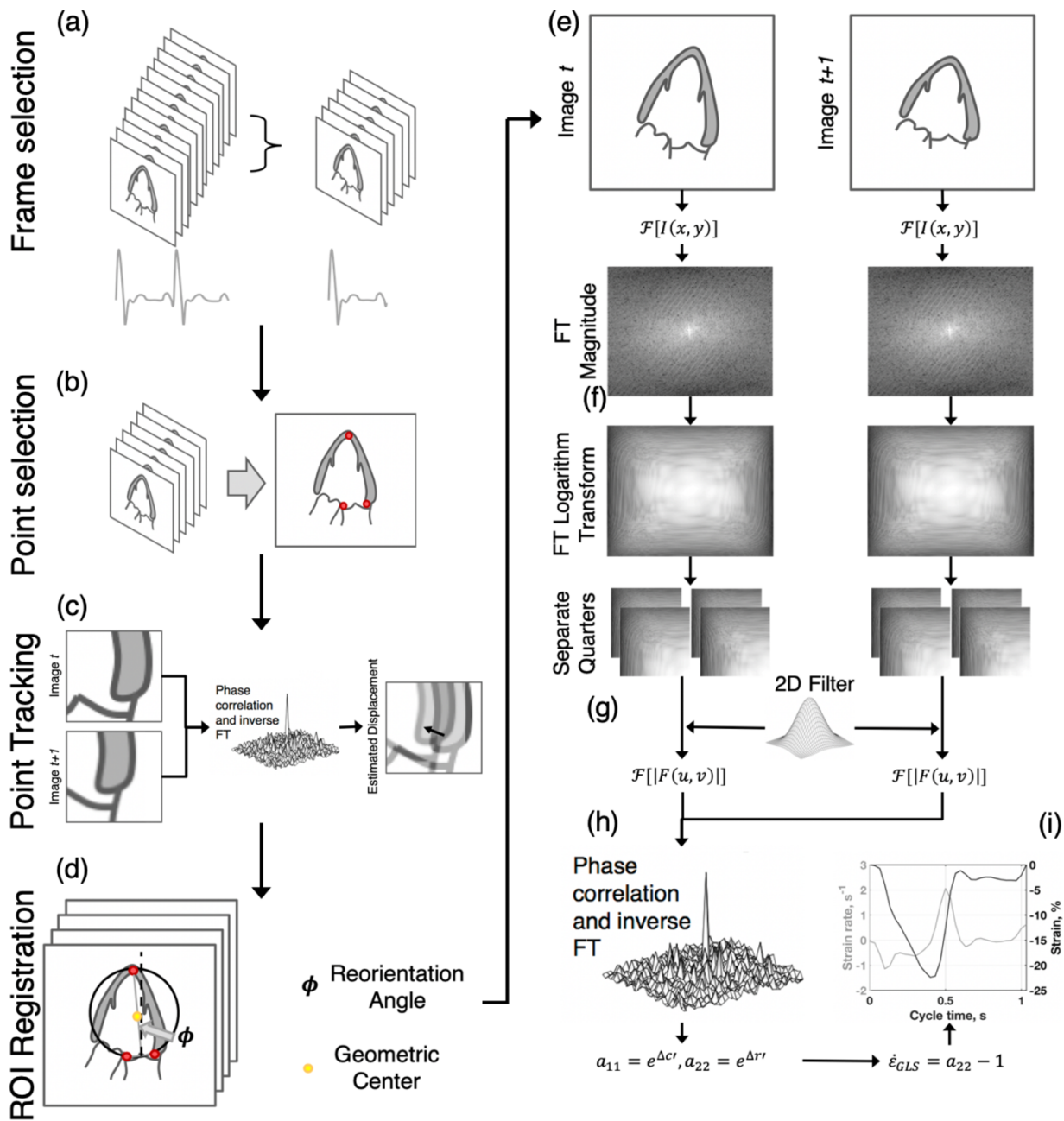

Figure 1: Illustration of the direct global longitudinal strain estimation algorithm. (a) Frames for a single beat are selected from an input echocardiogram. (b) User inputs for the apex and annulus positions from a reference frame are provided. (c) The inputs are tracked temporally. (d) Frame co-registration is performed. (e) The LV is cropped from each frame, and these sub-images are Fourier-transformed. (f) The FT magnitude is calculated, interpolated onto a logarithm-basis, and separated into four sub-images. (g) Each sub-image is Fourier-transformed and convolved with a phase filter. (h) Ensemble phase correlation is performed, producing a correlation plane with a peak shifted from the plane center. This shift corresponds to a frame pair strain rate. (i) Strain is computed by temporally integrating the strain rate estimates.

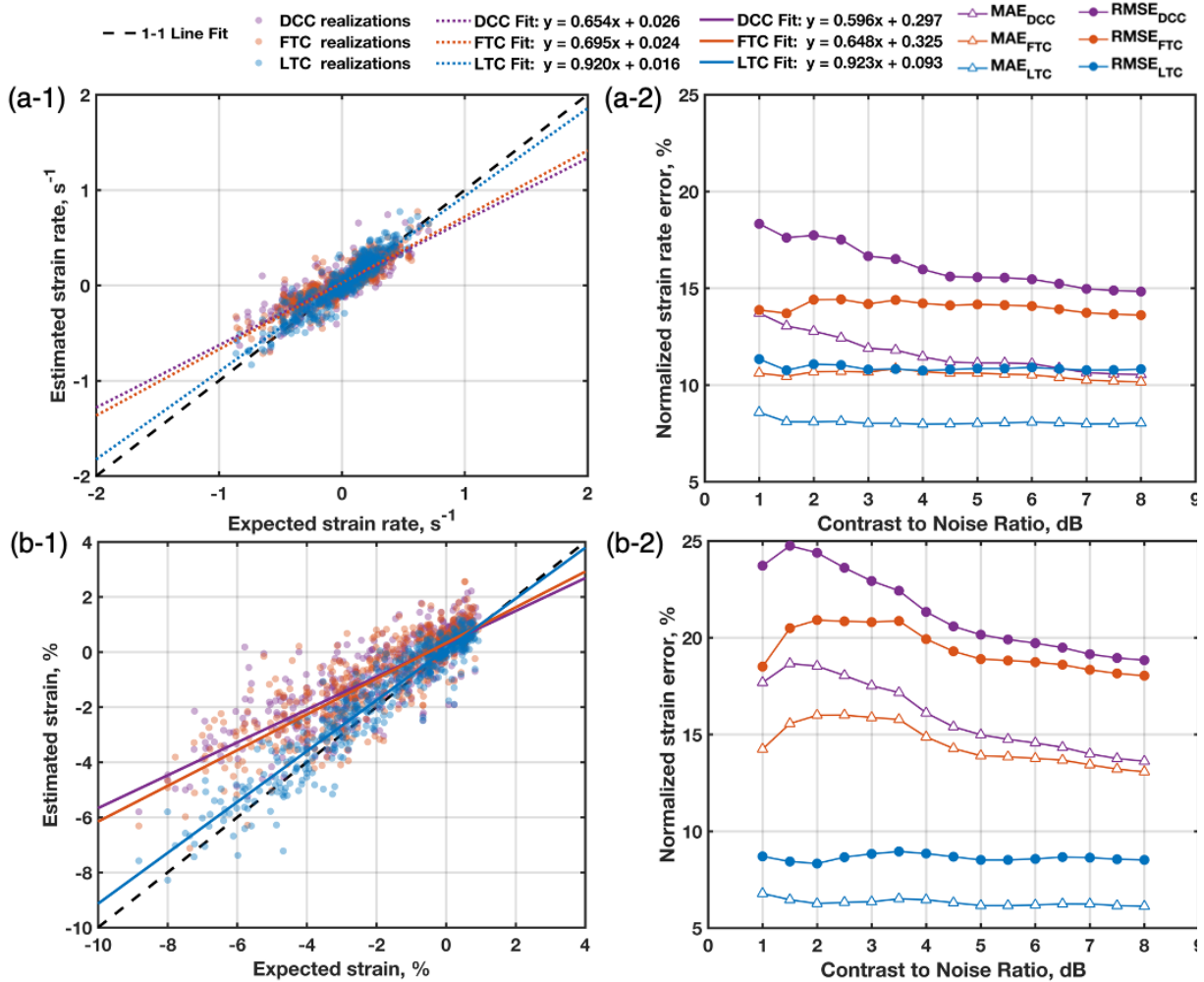

Figure 2: (Left) Direct comparison of measurements to ground truth values and (right) normalized mean absolute error (MAE) and root mean square error (RMSE) as a function of contrast-to-noise ratio (CNR) for (a) GLSr and (b) GLS quantities. Measurements were performed using the Direct Cross-Correlation method (DCC), Fourier Transform Correlation (FTC), and Fourier-based Logarithm Transform Correlation (LTC).

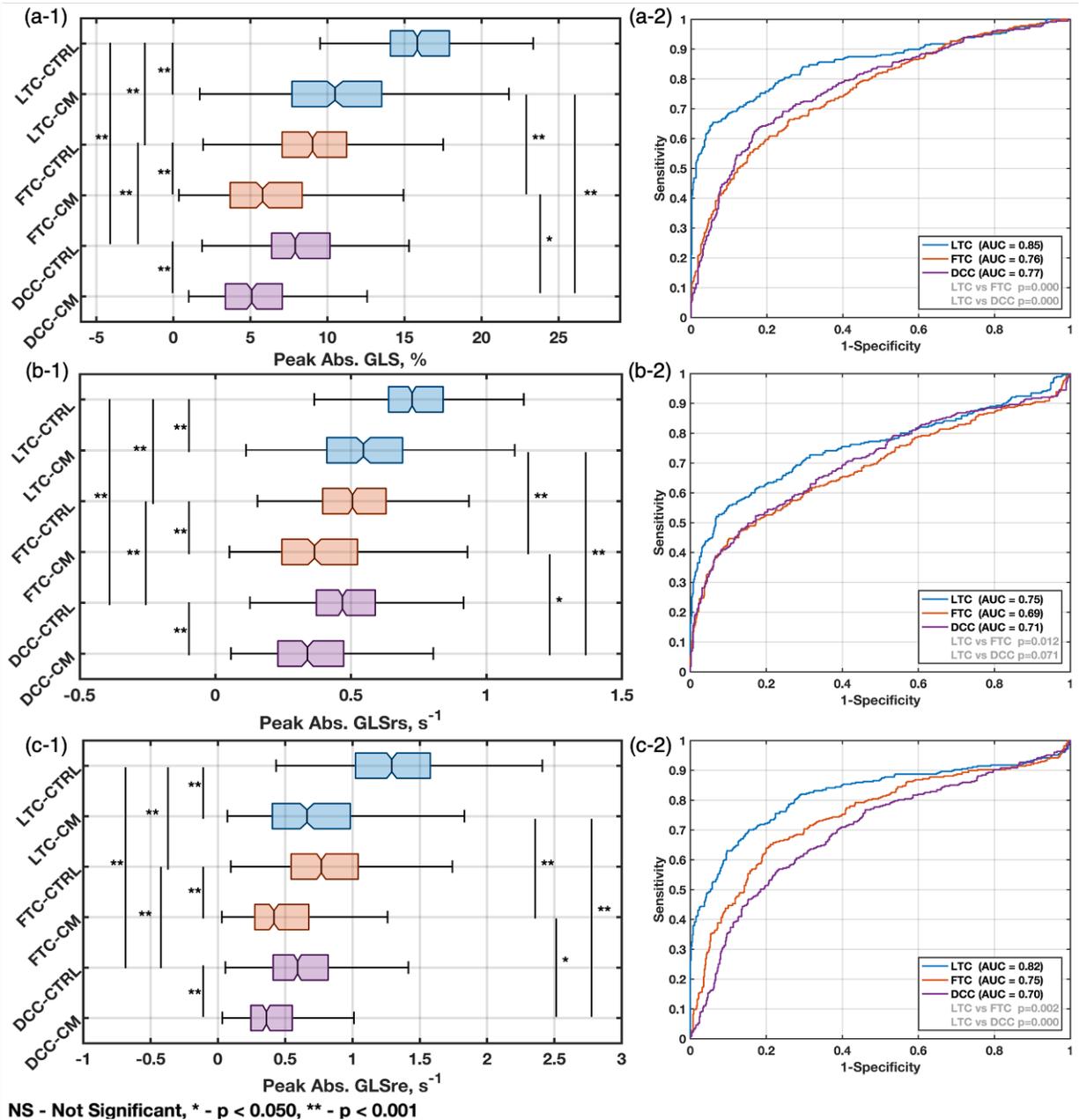

**NS - Not Significant, * - p < 0.050, ** - p < 0.001**

Figure 3: Distribution of measurements and significance tests for each GLS measurement method on observing (a-1) peak absolute GLS, (b-1) peak absolute systolic GLSr, and (c-1) peak absolute diastolic GLSr. (2) Receiver operating characteristic (ROC) curves displaying the ability for the LTC method estimated parameters to distinguish between normal and abnormal cardiac disease states based on (a-2) peak absolute GLS, (b-2) peak absolute systolic GLSr, and (c-2) peak absolute diastolic GLSr. The analysis was performed on a set of control subjects (CTRL) and subjects with cardiomyopathies (CM).

## Tables and Table Legends

Table 1: Demographics of the study cohort for each disease state.

| Characteristics | Control (*n = 33*) | DCM (*n = 4*) | HCM (*n = 16*) |
|---|---|---|---|
| Age (years) | 17.98 ± 8.86 | 14.50 ±6.24 | 18.74 ± 10.47 |
| BSA (m$^2$) | 1.66 ± 0.56 | 1.52 ±0.64 | 1.81 ± 0.69 |
| Height (cm) | 159.25 ± 29.11 | 147.90 ± 57.60 | 159.81 ± 30.86 |
| Weight (kg) | 63.50 ± 30.42 | 57.60 ± 35.65 | 75.71 ± 41.03 |
| Heart Rate (bpm) | 67.47 ±17.26 | 92.50 ± 33.81 | 72.88 ± 18.78 |

Table 2: Indices for LV dimensions and functional parameters.

| *Ventricular Dimensions* | Control (*n = 33*) | DCM (*n = 4*) | HCM (*n = 16*) |
|---|---|---|---|
| End Diastolic Volume (ml) | 98.85 ± 39.19 | 178.75 ± 83.92 | 96.28 ± 37.66 |
| End Systolic Volume (ml) | 37.80 ± 15.84 | 117.75 ± 60.31 | 36.03 ± 18.40 |
| Stroke Volume (ml) | 61.29 ± 24.39 | 61.00 ± 33.32 | 59.44 ± 21.24 |
| Ejection Fraction (%) | 62.16 ± 3.50 | 34.25 ± 14.93 | 63.06 ± 6.01 |
| *Functional Parameters* | | | |
| E-wave velocity (cm s$^{-1}$) | 82.20 ± 19.70 | 102.25 ± 29.80 | 83.19 ± 19.59 |
| A-wave velocity (cm s$^{-1}$) | 42.84 ± 10.95 | 61.25 ± 34.74 | 61.13 ± 33.65 |
| e' velocity (cm s$^{-1}$) | 17.58 ± 3.22 | 11.38 ± 2.63 | 10.51 ± 3.04 |
| E/A ratio | 2.00 ± 0.59 | 2.24 ± 1.62 | 1.55 ± 0.50 |
| E/e' ratio | 4.06 ± 1.21 | 8.22 ± 4.30 | 7.34 ± 2.58 |